# Molecular librations in a liquid investigated by ultrafast optical Kerr effect spectroscopy

V.G.Nikiforov and S.A.Moiseev

*Zavoisky Kazan physical –technical Institute of Russian Academy of Sciences,
Sibirsky trakt 10/7, Kazan, 420029, Russia
E-mail: vgnik@mail.ru, samoi@yandex.ru*

**Abstract**

We propose a detail theoretical approach modeling molecular librations in liquid investigated by ultrafast optically heterodyne detected optical Kerr effect (OHD-OKE). The approach has been applied for fluoroform ($CHF_3$) molecular liquid and new information about spectrum, relaxation time and time domain librations has been obtained.

**Introduction**

Ultrafast dynamics and interaction of molecules in a liquid phase are subjects of numerous investigations. In particular it is due to the increased interest in molecular mechanisms determining very sophisticated biological processes such as ferment reactions, matrix biosynthesis and others[1]. Interaction of a molecule with its nearest neighbors in a liquid leads to molecular librations in the local potentials formed by these interactions [2]. Typical time period of a molecular translational jump between the two different potentials is about $10^{-8}$-$10^{-10}$ s at room temperature. Orientational molecular diffusion has a shorter time period: $10^{-10}$-$10^{-12}$s. Therefore, experimental investigation of librational dynamics requires the technique, which is able to detect nonstationary signal with temporal duration not bigger than 100 fs. One of the coherent ultrafast laser spectroscopy methods, which can detect such a temporal librational response, is based on the using of nonlinear *optical Kerr effect* with *optically heterodyne detection* (OHD-OKE)[3]. Interpretation of the OHD-OKE experiments requires theoretical treatment of a fast third-order nonlinear optical response.

At the present stage the theoretical description of the optical nonlinear response in liquid remains a subject of topical researches [4-11]. The ultrafast laser pulses cover a broad range of the low-frequency transitions and give raise the different Raman active molecular motions, which can not be longer described as simple two-level systems. First general theoretical analysis of the fast nonlinear molecular response on the action of the femtosecond laser fields was given in the works of Tanimura and Mukamel, who proposed path-integral techniques in Liouville space[12,13]. Unfortunately the problem of molecular dynamics in a liquid investigated by high-order optical response complicates an application of such theories to analysis of experimental data. These difficulties lead the interpretations of OHD-OKE experimental data[14-18] are based on the phenomenological models[19-27] and using of the Fourier-transform method that separates the nuclear and non-resonant electronic contributions in the total nonlinear-optical transient response [28]. These investigations demonstrate that the OHD-OKE signal shows a significant deviation from single exponential decay of the rotational molecular response, which typically corresponds to the coherent librational response[24,29] at time delays smaller than 1 ps.

Traditionally the phenomenological description of the nonstationary librational signal uses models of overdamped and underdamped oscillators [22,23,30-32] which assume an influence of fast fluctuation forces described by a friction constant in the models. The models imply that the fluctuations are much faster than the molecular oscillations having period about tens femtosecond. In particular the interpretation of carbon disulfide $CS_2$ rotational molecular dynamics within the framework of these models gives the irreversible time of librational response, which is about ten times shorter than the oscillational periods[24,30,31] and thus the result does not rather correspond to a real physical picture of the molecular librations. In this work we propose an enriched phenomenological librational model by introducing an initially unknown distribution function of librational frequencies which, as we show below, can be easily found from the numerical analysis of the OHD-OKE data. We demonstrate that the fast decay of the coherent librations can be described in such model with a high accuracy without introducing the overly fast irreversible relaxation processes. With a purpose of the complete analysis of the OHD-OKE experiment data we also briefly analyze the responses of the intramolecular motions and the orientational diffusion without using Fourier-transform method. In the final part of this paper we apply the proposed approach for analysis of the fluoroform (CHF$_3$) OHD-OKE signal investigated by Laurent et. al.[33] and this analysis have allowed us to obtain the frequency distribution function, amplitude and relaxation time period of the coherent librations.

**Modeling the OHD-OKE signal**
We consider a traditional OHD-OKE experimental schema with a two pulse excitation of the medium.[3, 23] The first intensive pump pulse $E_{py}$ excites nonstationary anisotropy polarizability of a liquid probed by the second weak pulse $E_o$ with the time delay $\tau$. Let us assume that the electric field $E_{py}$ of the first light pulse is polarized along y-axis (see Fig.1)

$$E_{py}(\omega_o, t) = E_{py}(t)\cos(\omega_0 t), \tag{1}$$

where the optical field frequency $\omega_o$ belongs to the transparency window of the medium spectrum. The probe pulse $E_o$

$$E_o(\omega_o, \tau, t) = \sqrt{2} E_o(t-\tau)\cos(\omega_o(t-\tau)), \tag{2}$$

is polarized angularly 45° with the pump pulse polarization (see Fig.1). Changing of the initial probe pulse polarization by the nonstationary anisotropy medium modulates the probe pulse intensity in the output of crossed polarizers. The signal/noise ratio is considerably improved by using such optically heterodyne technique.[23, 34, 35] A heterodyne signal is formed in the mixing process of the probe pulse with the local oscillator field $E_{lo}$ polarized perpendicularly to the probe field polarization with additional 90° phase shift:

$$E_{lo}(\omega_o, \tau, t) = \sqrt{2} E_{lo}(t-\tau)\sin(\omega_o(t-\tau)). \tag{3}$$

The temporary envelopes of the electric fields in formulas (1) – (3) is given with Gaussian shape (where $\tau_l$ is a pulse duration):

$$E_{py}(t) \propto E_o(t) \propto E_{lo}(t) \propto \exp(-t^2/2\tau_l^2), \tag{4}$$

Let us introduce the molecular polarizability tensor $\widehat{P}_{mol}$ in the local molecular coordinate system: $\alpha_{xx} = \alpha_{zz} = \alpha_\perp$, $\alpha_{yy} = \alpha_\uparrow\uparrow$ with other components equal zero. We will describe y-direction of the molecular axis with azimuth angles $\theta$ and $\varphi$ in the laboratory coordinate system (see Fig. 1). In this case an OHD-OKE signal may look like (see Ap. 1):

$$S(\tau) = \frac{1}{T}\lim_{T\to\infty} \int_{-T/2}^{T/2} s(\tau, t)dt, \tag{5}$$

where $T$ is registration time, $s(\tau, t)$ is an instant response having the form:

$$s(\tau, t) \propto \{R_m(t) + R_{el}(t)\}E_{lo}(\tau, t)E_o(\tau, t) \tag{6}$$

where

$$R_{el}(t) = \frac{3}{2}\gamma I_{py}(t), \tag{7}$$

$$R_m(t) = (\alpha_{\uparrow\uparrow}(t) - \alpha_\perp(t))[\cos^2(\theta(t)) - \sin^2(\theta(t))\cos^2(\varphi(t))]. \tag{8}$$

The electronic response function $R_{el}(t)$ is proportional to the coefficient of instant electronic cubic huperpolarizability $\gamma$ and the response represents the temporary autocorrelational function of the pump, probe and local oscillator pulses. The molecular response function $R_m(t)$ is proportional to the molecular polarizability anisotropy $(\alpha_{\uparrow\uparrow}(t) - \alpha_\perp(t))$ modulated by the molecular rotational motions. For the convenience of calculation we introduce new variables

$$\alpha_{\uparrow\uparrow}(t) - \alpha_\perp(t) = \Delta\alpha + \delta\alpha(t), \tag{9}$$

where $\Delta\alpha$ is a constant parameter and $\delta\alpha(t)$ describes the influence of nuclear oscillations on the molecular polarizability. Taking into account the pump pulse polarization one should note that the pump pulse affects the molecular orientation angle $\theta$ while the angle $\varphi$ is not changed. Averaging the response (8) over the angle $\varphi$ we obtain:

$$R_m(\theta,t) = (\Delta\alpha + \delta\alpha(t))\frac{3}{2}\left(\cos^2(\theta(t)) - \frac{1}{3}\right). \tag{10}$$

Eqs. (5), (6), (7) and (10) describe the temporal behavior of the OHD-OKE signal through the function $R_m(t)$, which contains librational, orientational diffusive and intramolecular oscillation responses in the molecular liquid. A detail theoretical treatment of the molecular responses is given below.

**Intramolecular oscillations**
We describe the intramolecular nuclear oscillation response $R_n(t)$ using Eq. (10). Note that the interaction energy of a molecule with the pump pulse is less than $kT$ in traditional conditions of the OHD-OKE experiments. Therefore it is possible to ignore weak correlations between the rotational and intramolecular motions and to find the following response for intramolecular oscillations:

$$R_n(\theta_o,t) = R_m(t)\big|_{\theta(t)\equiv\theta_o} = (\Delta\alpha + \delta\alpha(\theta_o,t))\frac{3}{2}\left(\cos^2(\theta_o) - \frac{1}{3}\right), \tag{11}$$

where $\theta_o$ is an equilibrium angle of molecular orientation in the local potential $U$ of the nearest neighbors. Non-equilibrium polarizability $\delta\alpha(\theta_o,t)$ of all the intramolecular oscillations contains responses of the modes with normal molecular coordinates $q_i(\theta_o,t)$:

$$\delta\alpha(\theta_o,t) = \sum_{i=1}^{N}\alpha_i^1 q_i(\theta_o,t) \tag{12}$$

where $\alpha_i^1$ is a molecular polarizability coefficient of the i-th mode, $N$ is a number of oscillation modes. The coherent intramolecular dynamics in OHD-OKE experiments is described by the harmonic oscillator equation with the external force proportional to the pump field intensity[36] $I_{py}(t)$:

$$\frac{d^2}{dt^2}q_{ni}(\theta_o,t) + \frac{2}{\tau_{n_i}}\frac{d}{dt}q_{ni}(\theta_o,t) + \Omega_{ni}^2 q_{ni}(\theta_o,t) = \frac{\alpha_i^1}{2m_i} I_{py}(t)\cos(\theta_o), \qquad (13)$$

where $m_i$ is a reduced mass of the i-th oscillator mode, $\Omega_n$ and $\tau_n$ are the frequency and the relaxation time of the mode. The solution for the oscillation coordinate $q_i(\theta_o,t)$ is given as follows:

$$q_i(\theta_o,t) = \frac{\alpha_i^1}{2m_i}\cos(\theta_o)\Phi_n(\Omega_{ni},\tau_{ni},t), \qquad (14)$$

$$\Phi_n(\Omega_n,\tau_n,t) = (\Omega_n^2 - \tau_n^{-2})^{-1/2} \int_0^\infty I_{py}(t-t')\cdot\exp\left(-\frac{t'}{\tau_n}\right)\cdot\sin\left[(\Omega_n^2-\tau_n^{-2})^{1/2}\cdot t'\right]\cdot dt' \qquad (15)$$

After averaging Eq. (11) over the equilibrium angles $\theta_o$ we obtain the total intramolecular response:

$$R_n(t) \propto \sum_{i=1}^N \frac{\alpha_i^1}{m_i}\Phi_n(\Omega_{ni},\tau_{ni},t). \qquad (16)$$

The pump pulse excites most effectively Raman active oscillation modes having the period, which are comparable or larger than the pump pulse duration. Therefore the typical OHD-OKE experiment signal contains as a rule the coherent oscillation of "heavy" molecular fragments. Usually the excitations of coherent proton oscillations in molecules are not realized in typical OHD-OKE experiments, while slower oscillations of benzol ring and –OH, -CH$_3$, -Cl fragments can be detected with the pulse duration about 100 fs.[14, 37] In large molecules with the different masses of the fragments, a number of slow oscillating modes give contributions to the signal. It should be noted here that one can overcomes identification problem of the oscillation modes in the total OHD-OKE signal in particularly by an using of multi-pulse variants of the pumping in the OHD-OKE technique.[38]

**Orientational diffusion**
Laser pulses excite coherent rotational motions of molecules with anisotropy polarizability ($\Delta\alpha \neq 0$). Using Eq. (10) and taking into account $|\Delta\alpha| \gg |\delta\alpha(t)|$ we ignore the weak terms proportional to $\delta\alpha(t)$ while modeling orientational rotation. Frenkel's theory[2] includes two types of molecular rotations in liquid: librations and orientational diffusion. Orientational diffusion takes place due to rotational jumps between different equilibrium orientational angles of molecules. For small jump angles $\theta'_o - \theta_o \ll 1$ and $\varphi'_o - \varphi_o \ll 1$ this motion is modeled by the orientational diffusive equation[2, 39]. Experimental data[14, 23, 37] demonstrates that the relaxation time of orientational diffusion $\tau_d$ is tens or hundreds times as longer as the relaxation time of librations. Using Eq. (10) we describe orientational diffusion through the following function of the order parameter $Q_d(\theta) = \frac{3}{2}\cdot(\cos^2(\theta)-\frac{1}{3})$:

$$R_d(t) = \left\langle R_m(t)\big|_{|\delta\alpha|\ll|\Delta\alpha|}\right\rangle_{\theta_o} = \Delta\alpha\langle Q_d(\theta_o,t)\rangle_{\theta_o}, \qquad (17)$$

where $\langle ... \rangle_{\theta_o}$ is average procedure over all the equilibrium molecular orientations. Finally we obtain the orientational diffusive response $R_d(t)$ as follows (see Ap.2)

$$R_d(t) \propto \Delta\alpha D \int_{-\infty}^{t} \eta(t') \exp\left(\frac{t'-t}{\tau_d}\right) dt', \qquad (18)$$

where $\eta(t) = \Delta\alpha \cdot I_{py}(t)/kT$, $D=1/(6\tau_d)$ is a diffusive coefficient, $\tau_d$ is average time of a molecular orientational jump. The typical time of the orientational diffusive response is equal to several picoseconds for simple molecules[23] like $CS_2$ and increases by tens picoseconds for more complicated molecules like acetophenone and its derivatives.[14] The temporal response of orientational diffusion $R_d(t)$ has picosecond exponential decay after pump pulse excitation. This response gives long decay "tails" in OHD-OKE signals corresponding to extended or flat molecules.

**Molecular librations**
We consider librations of molecules allocated in the different local potentials determining the librational frequencies. A variety of the locally realized potentials leads to the spectral broadening of the librational frequencies (having width about 50 cm$^{-1}$ for acetonitrile[40]). We model the librations of a molecular ensemble using one-particle distribution function $f_{\theta_o}(\Omega_{lib}, \theta, t)$:

$$f_{\theta_o}(\Omega_{lib}, \theta, t) = \rho(\Omega_{lib}) \delta(\theta - (\theta_o - \Theta(t))), \qquad (19)$$

where $\rho(\Omega_{lib})$ is usually an unknown equilibrium distribution function, $\Theta(t)$ is a total deviation angle of a molecule from equilibrium orientation $\theta_o$ and $\Theta(t)$ contains both the pump pulse molecular response and the stochastic oscillations $\phi_{st}$. The molecular ensemble response $R_{lib}(\theta_o, t)$ is calculated through averaging over the orientational angle $\theta$, the stochastic oscillations $\phi_{st}$ is described by introducing of the distribution function of angle orientations $1/\sigma_{st}$ and librational frequencies $\Omega_{lib}$:

$$\begin{aligned} R_{lib}(\theta_o, t) &= \int_{-\pi}^{\pi} d\theta \int_{\sigma_{st}} \sigma_{st}^{-1} d\phi_{st} \int_{0}^{\infty} d\Omega_{lib} f_{\theta_o}(\theta, \Omega_{lib}, t) R_m(\theta, t) \\ &= \int_{\sigma_{st}} \sigma_{st}^{-1} d\phi_{st} \int_{0}^{\infty} d\Omega_{lib} \rho(\Omega_{lib}) R_m(\theta_o - \Theta(t), t), \end{aligned} \qquad (20)$$

where we use normalization $\int d\phi_{st}/\sigma_{st} = 1$ and symmetry demand $\int \phi_{st} d\phi_{st}/\sigma_{st} = 0$. In further calculations we take into account that $\Theta(t) \ll 1$ therefore the response $R_m(\theta_o - \Theta(t), t)$ takes the form:

$$R_m(\theta_o - \Theta(t), t) \cong R_m(\theta_o) + \delta R_m(\theta_o, t), \qquad (21)$$
$$\delta R_m(\theta_o, t) = \frac{3}{2} \Delta\alpha \Theta(t) \sin(2\theta_o).$$

Further substitution of Eq. (21) in Eq. (20) and after averaging over the equilibrium molecular orientations $\theta_o$ we obtain:

$$R_{lib}(t) = \langle R_{lib}(\theta_o,t) \rangle_{\theta_o}$$
$$= -\frac{3}{4}\Delta\alpha \int_0^\pi d\theta_o \int_{\sigma_{st}} \sigma_{st}^{-1} d\phi_{st} \int_0^\infty d\Omega_{lib}\, \rho(\Omega_{lib},t)\sin(\theta_o)\sin(2\theta_o)\Theta(\Omega_{lib},\phi_{st},\theta_o,t). \quad (22)$$

Thermal molecular motions induce fluctuations of the local potential $U(t) = U_o + \delta U(t)$ at time scale $t \sim 1$ ps with the correlation time $\tau_c \sim \Omega_{n,lib}^{-1}$, where $\Omega_{n,lib}$ is a librational spectral width. In pure liquids the fluctuation of librational frequency $\delta\Omega_{lib} = \hbar^{-1}\delta U$ affecting the molecular spectral diffusion is proportional to the librational amplitude. For the small fluctuation amplitudes, the condition $\delta\Omega_{lib} \ll \Omega_{n,lib}$ is satisfied and molecular coherent librations decay by molecular dephasing which takes place due to large spectral broadening ($t\Omega_{n,lib} \gg 1$, $t\delta\Omega_{lib} \ll 1$). Ignoring the small spectral diffusion we describe the total deviation angle $\Theta(t)$ through the following sum:

$$\Theta(t) = \phi_{st}(t) + \beta(\theta_o,t), \quad (23)$$

where the angle $\beta(\theta_o,t)$ corresponds to the coherent molecular librations induced by the impact interaction of molecules with the pump field $W = -\vec{p}\vec{E}_p = -\hat{P}_{lab}\vec{E}_p\vec{E}_p$. Taking into account the molecular tensor polarizability $\hat{P}_{mol}$ we obtain the force affecting on the molecule

$$F_\theta^y(\theta_o,t)\Big|_{|\Theta|\ll 1;} = -\frac{\partial W}{\partial \theta_o} \cong -\frac{1}{2}\Delta\alpha \sin(2\theta_o) I_{py}(\tau,t) \quad (24)$$

in accordance with the equation for the deviation angle $\beta(\theta_o,t)$

$$\frac{\partial^2}{\partial t^2}\beta(\theta_o,t) + \Omega_{lib}^2 \beta(\theta_o,t) = J^{-1} F_\theta^Y(\theta_o,t), \quad (25)$$

for parabolic intermolecular potential $U_o$ with molecular inertial moment $J$. The solution of Eq. (25) is

$$\beta(\Omega_{lib},\theta_o,t) = \frac{\Delta\alpha}{2J}\sin(2\theta_o)\Phi_{lib}(\Omega_{lib},t), \quad (26)$$

where function $\Phi_{lib}(\Omega_{lib},t)$ is defined by the equation:

$$\Phi_{lib}(\Omega_{lib},t) = \Omega_{lib}\int_0^\infty I_{py}(t-t')\cdot\sin[\Omega_{lib}\cdot t']\cdot dt'. \quad (27)$$

Note that the librational amplitude is proportional to $\sin(2\theta_o)$ while the amplitude of intramolecular oscillations is proportional to $\cos(2\theta_o)$. After substituting Eqs. (23), (26) for Eq.

(22) and averaging over the angles $\theta_o$ and $\phi_{st}(t)$ we obtain the following solution for the librational response:

$$R_{lib}(t) \propto \Delta\alpha^2 J^{-1} \int_0^\infty d\Omega_{lib} \rho(\Omega_{lib}) \Phi_{lib}(\Omega_{lib}, t). \tag{28}$$

According to Eq. (28) the librational response depends on the distribution function $\rho(\Omega_{lib})$ characterizing inhomogeneous broadening of molecular liquid librations. The spectral shape of the function $\rho(\Omega_{lib})$ can be obtained from the experimental OHD-OKE signal with the use of Eq.(28). Below we present such analysis for fluoroform molecule $CHF_3$.

**$CHF_3$ molecular librations**
The fluoroform ($CHF_3$) is a polar molecule having the axial symmetry polarizability tensor and relatively large dipole moment. Recently Laurent et.al.[33] have reported fast OHD-OKE signal of molecule $CHF_3$ presented in Fig. 2. In this figure we also present molecular contributions in the total experimental signal: $R_n(t), R_d(t)$ and $R_{lib}(t)$. Molecular relaxation times and frequencies are presented in Table 1. The dash-dot line on Fig.2 demonstrates the inertial nature of the librational response: a signal reaches its maximum at time delay about 100 fs and then decays completely after about 500 fs. We have numerically modeled librational response using Eq. (28) and found that the function $\rho(\Omega_{lib})$ is close to the distribution function of Maxwell type

$$\rho(\Omega_{lib}) \propto \Omega_{lib}^2 \exp\left\{-\frac{\Omega_{lib}^2}{2\Omega_{n,lib}^2}\right\} \tag{29}$$

with the spectral width $\Omega_{n,lib} = 58$ cm$^{-1}$. Accuracy of the total OHD-OKE signal modeling consists about 95%. The exponential time decay of the orientational diffusive response $R_d(t)$ of $CHF_3$ molecules equals to $\tau_d = 0.8$ ps which corresponds to the molecular lifetime in the local potential $U$ (see Tab.1). Thus the decay of librational response for time domain $\tau \ll 1$ ps is determined by dephasing of a spectral broadened oscillations. Thus the librational fluctuations $\delta\Omega_{lib}$ of $CHF_3$ molecules at room temperature are less than 1 cm$^{-1}$ and do not affect the librational response. Equating the librational energy to the thermal energy at room temperature $A^2 \Omega_{n,lib}^2 J \approx kT$, where $A$ is librational amplitude, one can obtain that the librational amplitude $A$ is less $10^o$. Small magnitude of librations coordinates well with the assumed model of harmonic oscillation. It is interesting to compare the above analysis with the interpretation presented by Laurent et.al.[33], where the ultrafast response of OHD-OKE signal is modeled by the sum of the two types of motions. The first motion is underdamed oscillation with fixed frequency 70 cm$^{-1}$ and relaxation time 110 fs, the second motion corresponds to the overdamped oscillator with relaxation time equals to 75 fs. Such these short relaxation times hardly correspond to the real irreversible relaxation processes in the molecular liquids. The observed relaxation processes should correspond to the librational dephasing and the irreversibility rather corresponds to rotational jumps time interval is about $\tau_d = 0.8$ ps. with the time interval between rotational jumps for irreversibility is about $\tau_d = 0.8$ ps. We assume that the proposed theoretical approach based on the librational distribution function $\rho(\Omega_{lib})$ gives a more detail physical explanation of the rapid librational response (other unknown parameters are not used here).

**Discussion and conclusion**

Time domain optical spectroscopy investigations of the liquid molecular dynamics demonstrate rotational molecular responses with ultrafast relaxation time is less than 1 ps. These responses show inertial character and are often referred to collision- or interaction-induced molecular motions[40, 41] realized as the collective molecular librations in the local potentials[22, 42]. Mechanisms of the OHD-OKE-signal decay are determined both by the manifold molecular motions and by the local intermolecular interactions in liquids. There are several phenomenological models[22, 23, 30-32] describing the coherent librations but mechanisms of ultrafast relaxation are not clear yet. The traditionally used phenomenological models of the overdamped and underdamped oscillators can not be physically correct for the modeling of the data obtained in the OHD-OKE experiments[33] since the fluctuations of the local intermolecular potentials can not be much faster than the intramolecular oscillation periods which are about tens or hundreds femtoseconds and they are comparable with librational relaxation times[14, 21-23, 35, 37]. Thus the influence of such spatially homogeneous fluctuations hardly corresponds to irreversible decay of the observable coherent librations in the OHD-OKE experiments.

In this paper we have developed a theoretical approach and demonstrated that ultrafast decay processes of librational response can be described by a fast decoherence (dephasing) due to their huge inhomogeneous spectral broadening $\rho(\Omega_{lib})$ of the molecular librations. We describe theoretically the total molecular OHD-OKE response, which contains the signals of intramolecular oscillations (16), orientational diffusion (18) and librational motion (28). The response function $R_{lib}(t)$ of the coherent librations has been obtained in a general form containing the distribution function $\rho(\Omega_{lib})$ characterizing the liquid oscillations in the picosecond time domain. The developed approach allows an easy indentification of the librational distribution function by the numerically calculation of OHD-OKE response in the studied in the studied molecular ($CHF_3$) liquid. We have found that the fluoroform librations are characterized by the distribution function $\rho(\Omega_{lib})$ which shape at room temperature is close to the Maxwell distribution (29) demonstrating a three-dimensional nature of spatial local inhomogeneity of the medium. Spectral width of the destribution $\Omega_{n,lib} = 58$ cm$^{-1}$ determines the time decay of the coherent librations (about 500 fs), which is less than the average time of molecular diffusive jumps between nearest local potentials (0.8 ps). Our estimations of the librational amplitude indicate that deviation angle of the fluoroform molecule from the equilibrium state is less than 10$^o$ and the molecular librations are close to the harmonic character.

We note that the spectral distribution $\rho(\Omega_{lib})$ is determined directly by the intermolecular interactions, therefore the distribution should be sensitive to the thermodynamic parameters of the medium[29, 43]. Knowledge of the spectral distributions $\rho(\Omega_{lib})$ is important for more deep insight into the molecular dynamics of similar molecular liquids. In particular, recently it has been reported that the fast part of OHD-OKE response in acetonitrile $CH_3CN$ decreases twice at the temperature range from 8 to 75 C[44]. To our opinion, such strong OHD-OKE signal sensitivity indicates a considerable transformation of the librational spectral distribution $\rho(\Omega_{lib})$. With a purpose of detail investigation of the distribution function $\rho(\Omega_{lib})$, three pulse OHD-OKE experiment can be proposed to the purpose of the detail investigation of the distribution function $\rho(\Omega_{lib})$. The reason is that two pump pulses excite fifth-order optical response[11-13, 45], which is sensitive to the dephasing mechanisms of the induced nuclear motions. We expect that the Raman echo motion type of the librational response can be observed by the third probe pulse which would make it possible to detail examine the molecular dephasing and irreversible relaxation processes in femto- and picosecond time domain.

This work was supported by the grants: Russian Science Support Foundation and Dynasty Foundation and ICFPM, Russian Foundation for Basic Research № 03-03-96214 p 2003 and grant NIOKR № 06-6.3-154/2002 Ф (06).

**Appendix 1**

We are interested in the change of the probe pulse field $E_o$ propagating along z-axis in the laboratory coordinate system (see Fig.1) with the equaled x- and y- components $E_{ox}=E_{oy}=E_o$ according to the experiment schema.[3] The polarization components of the nonlinear medium are given as follows:

$$P_x^m(\theta,\varphi,t) = A_x(\theta,\varphi)(E_o(\omega_o,t)+E_{lo}(\omega_o,t)), \quad (A.1.1)$$
$$P_y^m(\theta,\varphi,t) = A_y(\theta,\varphi)(E_o(\omega_o,t)-E_{lo}(\omega_o,t))+3\gamma(I_{py}(t)E_o(\omega_o,t)-I_{py}(t)E_{lo}(\omega_o,t)),$$

where $P_y^m$ component contains the term describing electronic cubic molecular hyperpolarizability proportional to the coefficient $\gamma$. $A_x(\theta,\varphi)$ and $A_y(\theta,\varphi)$ are components of the polarizability tensor:

$$A_x(\theta,\varphi) = \alpha_\perp\left(\cos^2(\varphi)\cos^2(\theta)+\sin^2(\varphi)-\frac{1}{2}\cos(\varphi)\sin(2\theta)\right)+$$
$$+\alpha_{\uparrow\uparrow}\left(\cos^2(\varphi)\sin^2(\theta)+\frac{1}{2}\cos(\varphi)\sin(2\theta)\right), \quad (A.1.2)$$
$$A_y(\theta,\varphi) = \alpha_\perp\left(\sin^2(\theta)-\frac{1}{2}\cos(\varphi)\sin(2\theta)\right)+\alpha_{\uparrow\uparrow}\left(\cos^2(\theta)+\frac{1}{2}\cos(\varphi)\sin(2\theta)\right).$$

Using the approximation of slowly varied amplitudes[39] one can obtain the expressions for the electric components $e_x$ and $e_y$ of light emitted by the optical thin medium:

$$e_x(\theta,\varphi,t) = \frac{1}{2}L\omega_o c^{-1}(E_o(t)\sin(\omega_o t)-E_{lo}(t)\cos(\omega_o t))\int_0^\pi\int_0^{2\pi}A_x(\theta,\varphi)\sin\theta\, d\theta\, d\varphi, \quad (A.1.3)$$

$$e_y(\theta,\varphi,t) = \frac{1}{2}L\omega_o c^{-1}(E_o(t)\sin(\omega_o t)+E_{lo}(t)\cos(\omega_o t))\int_0^\pi\int_0^{2\pi}\left(\frac{3}{2}\gamma I_{py}(t)+A_y(\theta,\varphi)\right)\sin\theta\, d\theta\, d\varphi,$$

where $L$ is the medium length. The field detected in the OHD-OKE experiment is $E'(\theta,\varphi,t) = \frac{1}{\sqrt{2}}(E_{lo}(t)+(e_x(\theta,\varphi,t)-e_y(\theta,\varphi,t)))$. We are interested in the response is proportional to the field $E_o(t)$:

$$s(t) \propto E_{lo}(t)E_o(t)\int_0^\pi\int_0^{2\pi}\left(A_y(\theta,\varphi,t)-A_x(\theta,\varphi,t)+\frac{3}{2}\gamma I_{py}(t)\right)\sin\theta\, d\theta\, d\varphi \quad (A.1.4)$$

The Eq. (A.1.4) describes the traditional OHD-OKE signal containing all nonstationary molecular responses.

**Appendix 2**

The interaction energy of a molecule with the pump pulse is written in the form:[39]

$$W = -\vec{p}\vec{E}_{py} = -\hat{P}_{lab}\vec{E}_{py}\vec{E}_{py} = -\left(\alpha_0 + \tfrac{2}{3}\cdot\Delta\alpha\cdot Q_d(t)\right)\cdot I_{py}, \qquad (A.2.1)$$

where $\alpha_0 = \tfrac{1}{3}(\alpha_{\uparrow\uparrow} + 2\alpha_\perp)$. Using the orientational diffusive model we obtain the following kinetic equation describing behavior of the parameter $\langle Q_d(t)\rangle$:[2,37,39]

$$\frac{1}{D}\cdot\frac{\partial\langle Q_d(t)\rangle}{\partial t} = -\frac{8}{3}\eta(t)\cdot\langle Q_d^2(t)\rangle + \left(\frac{4}{3}\eta(t) - 6\right)\cdot\langle Q_d(t)\rangle + \frac{4}{3}\eta(t) \qquad (A.2.2)$$

where $D=1/(6\tau_d)$ is the diffusive coefficient, $\eta(t) = \Delta\alpha\cdot I_{py}(t)/kT$. Taking into account the OHD-OKE experimental condition $\Delta\alpha\cdot I_{py} \ll kT$ we can simplify Eq. (A.2.2) and get the following

$$\frac{1}{D}\frac{\partial}{\partial t}\langle Q_d(t)\rangle = -6\langle Q_d(t)\rangle + \frac{4}{5}\eta(t,\tau). \qquad (A.2.3)$$

The solution of Eq.(A.2.3) is given in Eq. (18).

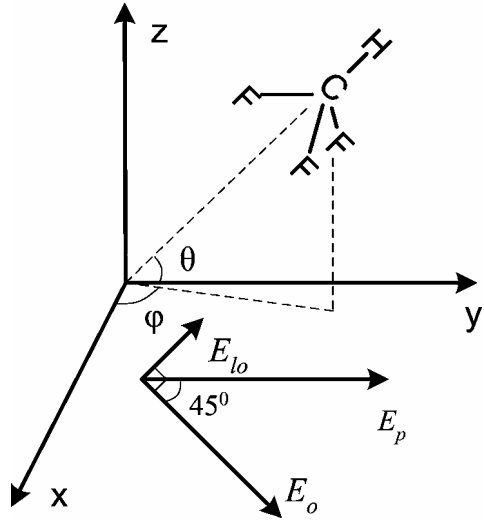

Figure 1. Molecular orientation and polarization vectors of pump pulse $E_{py}$, probe pulse $E_o$ and local oscillator field $E_{lo}$

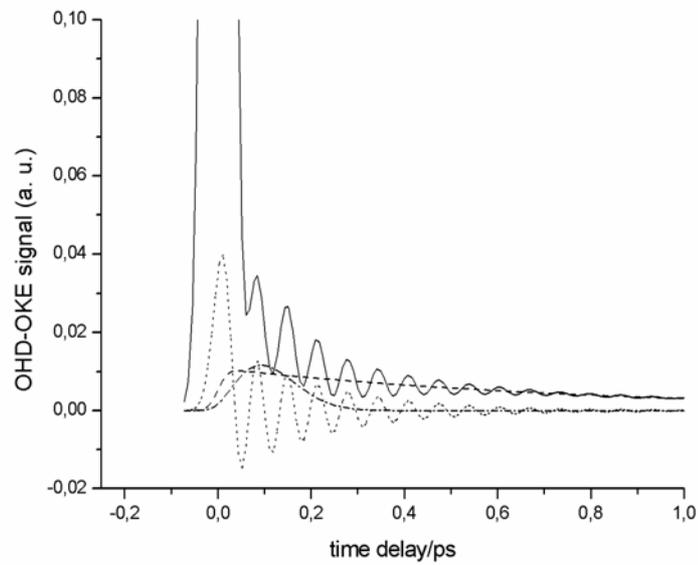

Figure 2. The total OHD-OKE signal and contributions of intramolecular response $R_n(t)$ (dot line), orientational diffusive response $R_d(t)$ (dash line) and librational response $R_{lib}(t)$ (dash-dot line)

Table 1. The modeling parameters of molecular dynamics $CHF_3$

| $\Omega_n$ | $\tau_n$ | $\tau_d$ | $\Omega_{n.lib}$ | $\tau_l$ |
|---|---|---|---|---|
| 512 cm$^{-1}$ | 0.2 ps | 0.8 ps | 58 cm$^{-1}$ | 24 fs |